\documentclass[aps,prl,twocolumn,superscriptaddress,amsmath,amssymb,longbibliography]{revtex4-1}
\usepackage{graphicx}
\usepackage{dcolumn,tabularx}
\usepackage{bm}
\usepackage{braket}

\usepackage{mathdots}
\usepackage[T1]{fontenc}
\usepackage[unicode=true]{hyperref}
\usepackage{booktabs}
\usepackage{mathtools}
\usepackage{color}
\usepackage{dsfont}
\usepackage{mathrsfs}

\begin{document}
	\renewcommand{\figurename}{Fig.}
	\title{Multipartite quantum resource distillation through local measurement programs}
	
	\author{Yan Wang}
	\affiliation{School of Physics, Hangzhou Normal University, Hangzhou 310036, China}
	
	\author{Shao-Qi Lin}
	\affiliation{School of Physics, Hangzhou Normal University, Hangzhou 310036, China}
	
	\author{Xi-Nuo Tao}
	\affiliation{School of Physics, Hangzhou Normal University, Hangzhou 310036, China}
	
	\author{Li-Jiong Shen}
	\affiliation{School of Physics, Hangzhou Normal University, Hangzhou 310036, China}
	
	\author{Yong-Nan Sun}
	\affiliation{School of Physics, Hangzhou Normal University, Hangzhou 310036, China}
	
	\author{Ze-Yan Hao}
	\email{hzy0526@mail.ustc.edu.cn}
	\affiliation{CAS Key Laboratory of Quantum Information, University of Science and Technology of China, Hefei 230026, China}
	\affiliation{Anhui Province Key Laboratory of Quantum Network, University of Science and Technology of China, Hefei 230026, China}
	\affiliation{CAS Center for Excellence in Quantum Information and Quantum Physics, University of Science and Technology of China, Hefei 230026, China}
	
	\author{Kai Sun}
	\affiliation{CAS Key Laboratory of Quantum Information, University of Science and Technology of China, Hefei 230026, China}
	\affiliation{Anhui Province Key Laboratory of Quantum Network, University of Science and Technology of China, Hefei 230026, China}
	\affiliation{CAS Center for Excellence in Quantum Information and Quantum Physics, University of Science and Technology of China, Hefei 230026, China}
	\affiliation{Hefei National Laboratory, University of Science and Technology of China, Hefei 230088, China}

	\author{Qi-Ping Su}
	\email{sqp@hznu.edu.cn}
	\affiliation{School of Physics, Hangzhou Normal University, Hangzhou 310036, China}
	
	\author{Chui-Ping Yang}
	\email{yangcp@hznu.edu.cn}
	\affiliation{School of Physics, Hangzhou Normal University, Hangzhou 310036, China}

	\begin{abstract}
		Distributed quantum resources in practical multi-user quantum networks are inevitably degraded by environmental noise, channel loss, and device-induced imperfections. 
		To address these issues, quantum resource distillation offers a fundamental approach to recovering stronger resources from imperfect states. 
		However, conventional implementations often require additional copies, dedicated physical filtering elements, or restrict to bipartite systems, posing challenges for scalable multipartite networks. 
		Here, we introduce the method of quantum resource distillation based on the local measurement program (LMP), which transfers completely positive maps into programmable measurement processes. 
		We experimentally demonstrate the performance of resource distillation through LMP in both bipartite and tripartite photonic systems, including the activation and enhancement of multipartite steering configurations. 
		To demonstrate the flexibility and extensibility of the LMP framework, we also show that virtual resource distillation can be naturally reformulated within it. 
		Our results establish a programmable and experimentally economical approach for distilling quantum resources in multipartite and higher-dimensional systems, thereby providing a practical route toward scalable quantum networks.
	\end{abstract}
	
	\maketitle
	
	\section{Introduction.}
	Quantum resources shared across distant nodes are essential for quantum information science \cite{RevModPhys.95.011003} and drive genuine advantages in quantum information processing, e.g., quantum teleportation \cite{PhysRevLett.70.1895,hu2023progress}, quantum-enhanced sensing \cite{RevModPhys.89.035002}, long-distance and secure communication \cite{RevModPhys.95.045006}, and scalable quantum computing \cite{RevModPhys.94.015004}. 
	In realistic scenarios, quantum resources are inevitably exposed to environmental disturbances that induce decoherence \cite{lidar,RevModPhys.95.045006}. Beyond merely degrading entanglement \cite{horodecki2009quantum}, such noise can also alter directional quantum correlations like quantum steering \cite{RevModPhys.92.015001}, thereby reshaping which parties still share usable quantum resources. 
	This motivates experimentally feasible methods capable of recovering and purifying useful quantum correlations  \cite{PhysRevLett.76.722,PhysRevLett.77.2818,gottesman1999demonstrating,pan2003experimental,t8qx-hvz7,zhang2025entanglement}, paving the way for more reliable quantum technologies  \cite{o2009photonic,PhysRevX.12.041023,Wendin_2017,RevModPhys.83.33,Azuma2015AllPhotonicRepeater,Wehner2018QuantumInternet,main2025distributed}.
	
	Quantum resource distillation provides a route to recovering stronger nonclassical correlations from imperfect resources \cite{RevModPhys.91.025001}. 
	The standard multi-copy protocols rely on additional noisy pairs and use entangling operations
	to postselect fewer outputs with enhanced entanglement \cite{PhysRevLett.126.010503,kbw2-fdqn}, but these requirements become increasingly demanding when scaling to multipartite or high-dimensional systems \cite{glc7-xy8t}. 
	More broadly, the distillation protocol can also be formulated as a probabilistic state transformation that enhances the operationally useful correlations of an imperfect resource \cite{PhysRevLett.128.110505}. 
	Within this perspective, local filtering implements a trace-nonincreasing and  completely positive (CP) map on the target state and retains only successful events \cite{gisin1996hidden,kwiat2001experimental,zhang:24,PhysRevLett.134.150201,PhysRevA.99.030101,Dominy2016GeneralFrameworkCP}. 
	Such filtering transformations can reveal hidden nonclassical correlations and have been extended from entanglement to steering distillation  \cite{PhysRevLett.124.120402,ku2022complete,liu2022distillation,PhysRevA.109.022411}. However, physical local filters are usually implemented by inserted, often fixed, optical elements or specific operations, so each target CP map requires a corresponding hardware configuration and introduces additional loss, alignment sensitivity, and calibration overhead \cite{kwiat2001experimental,zhang:24,PhysRevLett.134.150201}.
	
	In parallel, virtual resource distillation has been proposed and experimentally demonstrated to reproduce the statistics of target resource transformations without physically preparing the distilled state \cite{PhysRevA.109.022403,PhysRevLett.132.050203,PhysRevLett.132.180201}. It offers a different route, yet the existing demonstrations remain formulated as specific statistical reconstructions rather than as a general programmable measurement implementation of local CP maps \cite{PhysRevLett.132.180201}. 
	Moreover, recent work on temporal asymmetry in entanglement distillation \cite{glc7-xy8t} has shown that the choice and ordering of local operations can fundamentally determine whether distilled resources surpass classical thresholds in downstream protocols such as quantum teleportation, .underscoring the need for versatile control over the choice and ordering of local operations. 
	Therefore, a unified and programmable framework is still lacking for transferring local CP maps to local measurement procedures, so as to reproduce the same postselected distillation statistics without physically inserting the corresponding state transformation modules. 

	Here we develop the local measurement program (LMP) for quantum resource distillation, which implements the effect of local CP maps through programmable measurements rather than dedicated entangling gates or physical local filters. 
	We experimentally demonstrate the LMP framework in both bipartite and tripartite photonic systems under asymmetric noise, achieving distillation of both entanglement and steering, and revealing the hierarchical and directional transformation of tripartite steering correlations.  
	We also show that recently proposed virtual resource distillation can be reformulated within the same LMP picture, establishing a flexible route to resource distillation in multipartite and high-dimensional quantum systems. 
	Our method reduces experimental complexity and system errors, making resource distillation naturally compatible with scalable, multi-user quantum network architectures.

	\begin{figure}[t]
		\includegraphics[width=0.45\textwidth]{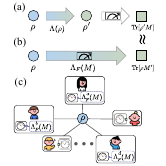}
		\caption{\textbf{Schematic illustration.} (a) The local filtering distillation is constructed through performing local operations $\Lambda(\rho)$ on the initial state $\rho$. (b) Local measurement program (LMP) distillation implements the filtering map on the measurement basis in $\Lambda_F(M)$. (c) LMP distillation in multipartite systems with  local operations selectively performing on i-th partite $\Lambda_F^{i}(M)$ ($i\in\{a,b,d\}$).
		}\label{th}
	\end{figure}
	
	\section{Theoretical framework.}
	As for the local filtering distillation in Fig.~\ref{th}(a), the transformation is typically represented by a CP, trace-nonincreasing map $\Lambda$ that acts on the input state $\rho$,
	with the postselected output state given by
	\begin{equation}
		\rho^\prime=\Lambda(\rho)=\frac{F\rho F^\dagger}{\mathrm{Tr}[F\rho F^\dagger]}, 
		\label{lf}
	\end{equation}
	where $F=\bigotimes_{j\in \mathcal{S}}F_j$ with $F_j$ acting on the $j$th local subsystem, and the resulting state $\rho'$ exhibits enhanced quantum resources (e.g., entanglement and steerability) \cite{gisin1996hidden,kwiat2001experimental,zhang:24,PhysRevLett.134.150201}.
	Such a map realizes quantum resource distillation by filtering out unwanted components locally, with a success probability given by $\mathrm{Tr}(F \rho F^\dagger)$. 
	The corresponding measurement probability is
	\begin{equation}
		\mathrm{Tr}\!\left[M_k\rho_F\right]
		=
		\frac{\mathrm{Tr}\!\left[M_k F\rho F^\dagger\right]}
		{\mathrm{Tr}\!\left[F\rho F^\dagger\right]}
		\label{dual_relation}. 
	\end{equation} 
	However, physically implementing local filters in Eq.~\eqref{lf} requires fixed optical elements that must be fabricated and aligned with high precision, making the protocol inflexible and increasingly demanding in multipartite or high-dimensional systems. 
	
	Motivated by these constraints, we construct the local measurement program (LMP) to distill quantum resources by equivalently transferring the action of the filtering map from the quantum state to the measurement operators, as illustrated in Fig.~\ref{th}(b). 
	Instead of physically applying $F$ to the quantum state, one can equivalently transform the measurement effects via the Heisenberg-picture dual map \cite{p7xt-s9nz}
	\begin{equation}
		\tilde M_k\;\Rightarrow\;  \Lambda_F^\dagger(M_k)=F^\dagger M_k F,
		\label{nlf}
	\end{equation}
	with the normalized probability $p(k|\rho_F)
	=
	\frac{\mathrm{Tr}(\rho \tilde M_k)}
	{\sum_j\mathrm{Tr}(\rho \tilde M_j)}
	=
	\frac{\mathrm{Tr}(\rho F^\dagger M_kF)}
	{\mathrm{Tr}(\rho F^\dagger F)}$.
	This establishes the central principle of the LMP: embedding the filter operation into the measurement procedure. More details of the LMP operation are provided in Sec.~S1 of the Supplementary Material (SM) \cite{SM}.
	
	More generally, for multipartite or higher-dimensional systems, the LMP protocol can be applied locally to each subsystem by transferring the corresponding local filtering operation to the measurement stage. When the local filter is selectively applied to the $i$th subsystem, as illustrated in Fig.~\ref{th}(c), its effect is absorbed into the corresponding local measurement via Eq.~\eqref{nlf}, while the measurement operators on all other subsystems remain unchanged.
	Compared with approaches acting directly on quantum states, the LMP avoids physically inserting and aligning multiple filtering elements across different subsystems, substantially simplifying the experimental apparatus.
	This programmable flexibility becomes essential when the ordering and selection of local operations fundamentally determine whether distilled resources can surpass classical thresholds in quantum information tasks \cite{glc7-xy8t}, highlighting a key scalability and feasible advantage of the LMP for complex quantum systems.

	\begin{figure}[t]
		\includegraphics[width=0.5\textwidth]{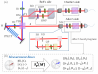}
		\caption{\textbf{Illustration of experiment.} (a) Pairs of entangled photons ($\ket{\Phi(\theta)_{b}}$) are generated via parametric down-conversion in a periodically poled potassium titanyl phosphate (PPKTP) crystal, with one output sent to beam displacer (BD) to further prepare tripartite states ($\ket{\Phi(\theta)_{t}}$). The noise module (NM) placed on Alice and Charlie's sides are detailed in the blue box. Three dotted boxes are the measurement apparatus including quarter-wave plate (QWP), half-wave plates (HWP), a BD or a polarized beam splitter (PBS). DM: dichroic mirror, ND: neutral density filter, M: mirror, BS: beam displacer, QP: quartz plate, IF: interference filters. (b) The concept of Alice's local program is to rotate the measurement bases from $\ket{0}$, $\ket{1}$, $\ket{+}$, $\ket{-}$, $\ket{+_i}$, $\ket{-_i}$ to $\ket{0_p}$, $\ket{1_p}$, $\ket{+_p}$, $\ket{-_p}$, $\ket{+_{ip}}$, $\ket{-_{ip}}$ along with the normalized coefficients $D_0$, $D_1$, $D_+$, $D_-$, $D_{+i}$ and $D_{-i}$, according to the operator $\mathcal{F}_a$ in \eqref{filter_operator}.
		}\label{expe}
	\end{figure}
	
	\section{Experimental setup.}
	The illustration of our experimental setup is depicted in Fig. \ref{expe}(a). The bipartite state is generated by a bi-photon entanglement source in the form of $\ket{\Phi(\theta)_{b}}=\cos\theta\ket{H_sH_i}+\sin\theta\ket{V_sV_i}$ \cite{Fedrizzi:07,PhysRevLett.118.140404}, where the subscript $s\ (i)$ represents the signal (idler) photon, and $H$ ($V$) is the horizontal (vertical) polarization. The parameter $\theta$ can be tuned by the half-wave plate (HWP) H1 in Fig. \ref{expe}(a). 
	For the tripartite case, the computational bases are encoded in photons' polarization and path degrees of freedom as the form:  $\ket{\Phi(\theta)_{t}}=\cos\theta\ket{H_sH_iu}+\sin\theta\ket{V_sV_il}=\cos\theta\ket{000}+\sin\theta\ket{111}$, where polarization information $H$ ($V$) and path modes $u$ ($l$) are encoded $0$ ($1$) \cite{cavalcanti2015detection}. 
	The local noise module (NM) with tunable strength is further inserted via the unbalanced interferometer, resulting in the bipartite and tripartite states as 
	\begin{equation}
		\rho_{b}=\eta\ket{\Phi(\theta)_{b}}\bra{\Phi(\theta)_{b}}+\frac{(1-\eta)}{2}\openone_A\otimes\rho_{C}^\theta,
	\end{equation}
	and 
	\begin{equation}
		\rho_{t}=\eta\ket{\Phi(\theta)_{t}}\bra{\Phi(\theta)_{t}}+\frac{(1-\eta)}{2}\openone_A\otimes\rho_{BC}^\theta,
		\label{rhoabc}
	\end{equation}
	with the ratio ($\eta$) controlled by the neutral density filters (NDs), $\rho_C^\theta=\mathrm{Tr}_A[\ket{\Phi(\theta)_{b}}\bra{\Phi(\theta)_{b}}]$ and $\rho_{BC}^\theta=\mathrm{Tr}_A[\ket{\Phi(\theta)_{t}}\bra{\Phi(\theta)_{t}}]$. 
	The measurement process consists of two polarization analyzers (on Alice and Charlie's sides) in the bipartite case, while an additional path analyzer (on Bob's side) is introduced in the tripartite case. Photons are collected by the single-photon detector after an interference filter with a 3-nm bandwidth and sent the output signal to a further coincidence device. We then perform the standard quantum state tomography for each experimentally prepared state \cite{PhysRevA.64.052312}, with the average fidelity for all bipartite and tripartite states $0.972(4)$ and $0.963(6)$. More experimental details are in S2 of the SM \cite{SM}. 
	
	The distillation process via LMP then works on the measurement device with the schematic principle shown in the dashed box of Fig. \ref{expe}(b). Here, we illustrate the LMP through an effective local filtering operation on Alice's side as 
	\begin{equation}
		F_a =
		\begin{pmatrix}
			\sin\theta & 0\\
			0 & \cos\theta
		\end{pmatrix}, 
		\label{filter_operator}
	\end{equation} 
	which is equivalent, up to an irrelevant overall normalization factor, to the inverse Schmidt-coefficient filter and satisfies \(F_a^\dagger F_a\le \mathbb I\). 
	Moreover, the form of $F_a$ is explicitly determined by the state parameter $\theta$, while independent of $\eta$. 
	The initial measurement bases (before applying LMP) are the eigenstates of three Pauli matrices, respectively represented in the basis as $\ket{0 (1)}=\ket{H (V)}$, $\ket{+ (-)}=\ket{H}\pm\ket{V}$ and $\ket{+_i (-_i)}=\ket{H}\pm i\ket{V}$.
	Alice's LMP will update the measurement bases according to  Eq. (\ref{nlf}), resulting the new measurement bases (after applying LMP) as $\ket{0_p (1_p)}=\ket{H (V)}$, $\ket{+_p (-_p)}=\sin\theta\ket{H}\pm\cos\theta\ket{V}$ and $\ket{+_{ip} (-_{ip})}=\sin\theta\ket{H}\pm i\cos\theta\ket{V}$ along with the normalized coefficients $D_{k}$ ($k\in\{0,1,+,-,+_i,-_i\}$). 
	We then utilize the local program on Alice's side, which can replace the physically optical components required from local operations by employing the basis rotation and subsequent data processing. The local program is also feasible and extendable to other parties, i.e., Bob or Charlie's sides. 
	
	\begin{figure}[t]
		\includegraphics[width=0.5\textwidth]{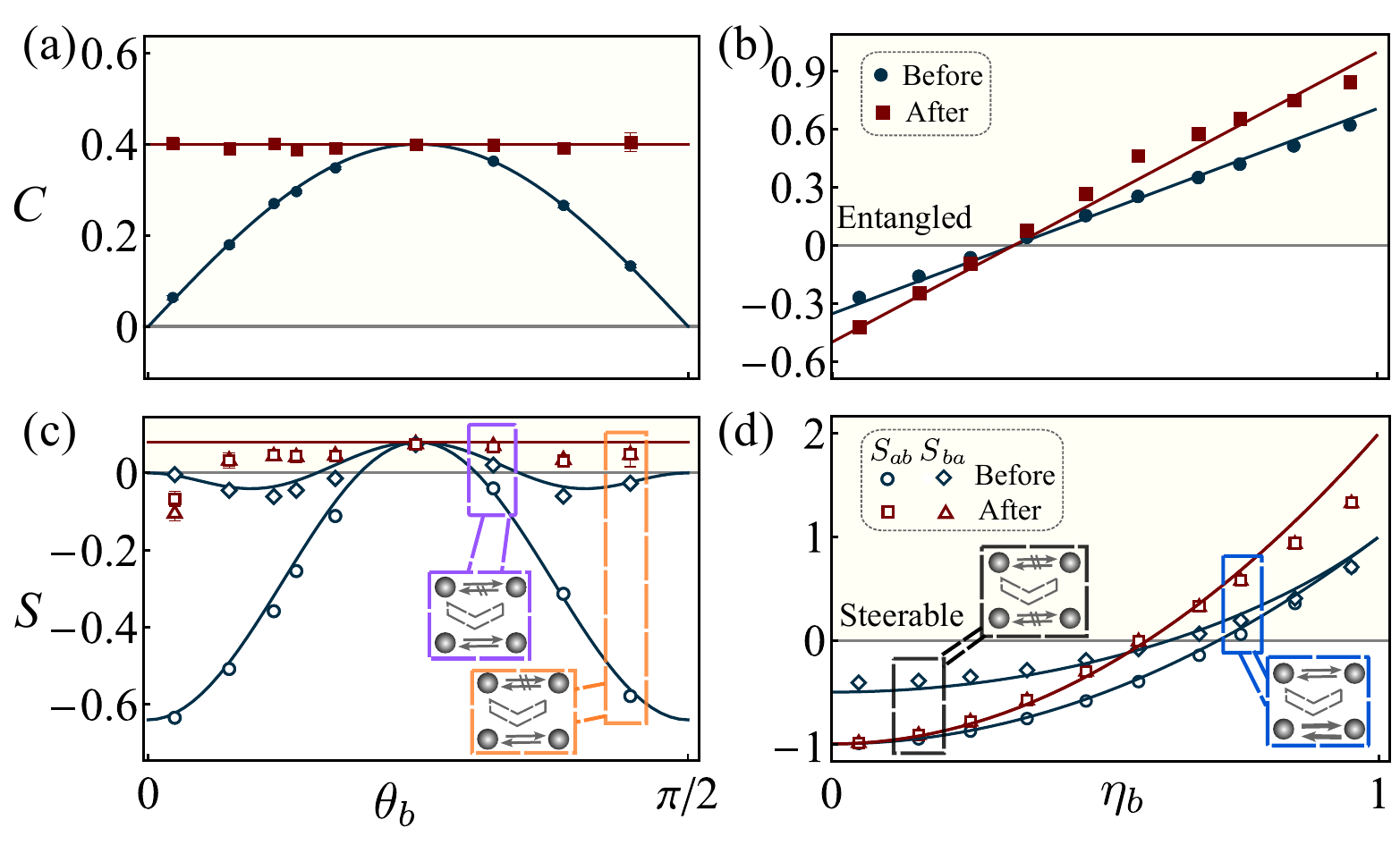}
		\caption{\textbf{Experimental results in the bipartite case.} The blue circles and rhombuses (red squares and triangles) represent before (after) distillation process. The solid and hollow markers respectively represent the concurrence ($C$) and steerability ($S$) of the bipartite states, where $S_{ab}$ ($S_{ba}$) are the steerability from Alice to Bob (Bob to Alice). The solid lines are the corresponding theoretical analysis. The yellow regions are denoted the entangled regions in (a) and (b), and the steerable regions in (c) and (d). The dashed boxes described the steering transformations. All errors are calculated assuming Poissonian statistics. }\label{res1}
	\end{figure}
	
	\section{Experimental results.}
	We first prepare series of bipartite states through subjecting to fixed asymmetric noise $\eta_{b}=40\%$ with varying $\theta_b\in(0,\pi/2)$, and fixed $\theta_{b}=\pi/8$ with varying asymmetric noise $\eta_b\in(0,1)$. The LMP is further applied to realize the distillation process with the experimental results denoted as solid and hollow markers representing the degree of entanglement ($C$) and steerability ($S$), which is demonstrated by $C > 0$ \cite{PhysRevLett.80.2245} and $S > 0$ \cite{PhysRevA.93.012108}. Here, the steering parameter $S$ is quantified from two directions, where $S_{ab}$ denotes steerability from Alice to Bob and $S_{ba}$ means Bob to Alice. The detailed calculations of the entanglement and steering criteria are in the S3 of SM \cite{SM}. The distillation performance of LMP in bipartite system is shown in Fig. \ref{res1}. 
	
	To obtain updated measurement bases and their normalization factors, the LMP adapts to different values of the parameter $\theta$ under a fixed noise module by updating the corresponding measurement bases and normalization factors. For states with fixed $\theta$, the same measurement bases and normalization are used for different noise ratios.  
	Owing to the hierarchical structures of quantum steering, we observe three distinct distillation scenarios, highlighted by the dashed boxes in Fig. \ref{res1}, i.e., the enhancement of initially two-way steerable states (the blue box), the conversion of one-way to two-way steerable states (the purple box), and the activation of previously unsteerable to two-way steerable states (the orange box). Moreover, the distillation effect of LMP would become ineffective (in black box) when varying parameter $\eta$ in (b) and (d), primarily because the mathematical form of the local filter employed in Eq. \eqref{filter_operator} fails to purify the system state unless the parameter $\eta$ exceeds a critical threshold. In this situation, only the initial states containing entanglement and steerability can be successfully distilled. Meanwhile, the observation that quantum steering requires higher threshold compared with that of entanglement reveals the hierarchical relationship between these quantum resources.

\begin{figure}[t]
		\includegraphics[width=0.5\textwidth]{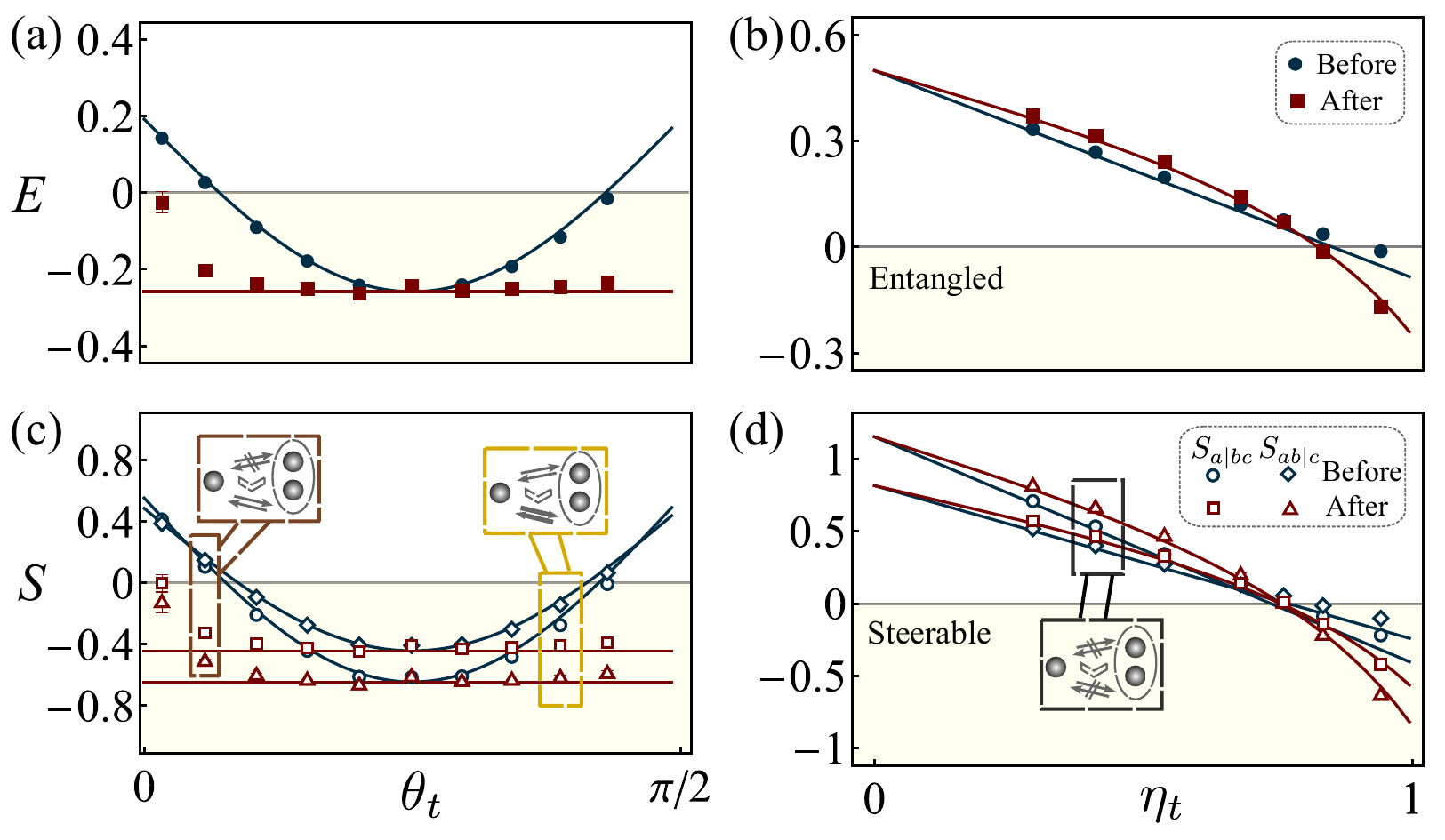}
		\caption{\textbf{Experimental results in the tripartite case.} The blue circles and rhombuses (red squares and triangles) represent experimental results before (after) distillation, and the solid and hollow markers respectively represent the entanglement ($E$) and steering witness ($S$), along with the theoretical analysis in solid lines. The parameters $S_{a|bc}$ ($S_{ab|c}$) in (c) and (d) denote the steering witness from Alice to Bob and Charlie (Alice and Bob to Charlie), where the steerability transformations are denoted in the dashed boxes. The yellow background is corresponding to entangled regions in (a) and (b), and steerable regions in (c) and (d). All errors are calculated under Poissonian statistics.
		}\label{res2}
	\end{figure}

	We then characterize the distillation effect of LMP in the tripartite system in Fig. \ref{res2}, including fixed asymmetric noise $\eta_{t}=10\%$ with varying $\theta_t\in(0,\pi/2)$ and fixed $\theta_{t}=\pi/9$ with varying asymmetric noise $\eta_t\in(0,1)$. In the tripartite case, the entanglement (steering) is verified with $W (S) < 0$ \cite{PhysRevLett.87.040401,cavalcanti2015detection}. Here, the tripartite quantum steering $S$ is calculated in both  one-sided ($S_{a|bc}$) and two-sided ($S_{ab|c}$) device independent scenarios. The detailed calculations of the entanglement and steering criteria are in S4 of the SM \cite{SM}. Similarly, the distillation protocol works across different values of $\theta_t$, while for fixed $\theta_t$ and varying $\eta_t$, successful distillation is observed only when the noise ratio remains below a threshold.
	In particular, we observe different configurations of multipartite steering distillation, including the enhancement and activation of steerability in the yellow and brown boxes from both one-sided and two-sided device independent scenarios. 
	The black box, by contrast, denotes the persistence of unsteerable states after the LMP processing.  
	These observations demonstrate that the LMP can flexibly manipulate multipartite steering configurations and be generalized to multipartite systems. 
	
	Our experimental results in both bipartite and tripartite systems show agreement with the theoretical predictions, confirming the effectiveness of the LMP in purifying quantum resources and providing a feasible approach for distilling genuine multipartite entanglement and quantum steering.  The deviations between experimental results and the theoretical analysis primarily stems from fluctuations of the entangled source, small calibration errors in the wave plates and the imperfections of our measurement apparatus. Notably, as the initial quantum resource decreases, i.e., the parameter $\theta$ approaches $0$ or $\pi/2$, the beforementioned experimental errors are amplified, making the distillation process more sensitive to such experimental errors and thereby degrading the distillation performance.

	Moreover, we reformulate the recently proposed
	virtual resource distillation \cite{PhysRevLett.132.050203,PhysRevLett.132.180201} in terms of the LMP framework, where a noisy Werner state is virtually distilled through a linear combination of measurement statistics. The desired virtual transformation can be decomposed into CP branches, $\sum_i \Lambda_i(\Psi^-)=\rho_\eta=(\mathbb{I}_4-\ket{\Psi^-}\!\bra{\Psi^-})/3$, where $\Lambda_1(\Psi^-)=\Psi^+/3$, $\Lambda_2(\Psi^-)=\rho_{00}/3$, and $\Lambda_3(\Psi^-)=\rho_{11}/3$ \cite{PhysRevLett.132.180201}. Instead of physically implementing these branches to the quantum state, LMP applies their dual action to the measurement operators, and the local state $\ket{\phi_\mu}$ after applying local operator $F_{i,\mu}$ is programmed as $F_{i,\mu}^{\dagger}\ket{\phi_\mu}$ after normalization, with the resulting programmed bases and coefficients for all three branches are summarized in Table~\ref{TA}. For example, in the $\Lambda_3$ branch, the local operator on Bob's side is $F_{3,B}=\ket{1}\!\bra{1}$, so an original measurement state such as $\ket{R}$ is programmed as $F_{3,B}^{\dagger}\ket{R}\propto\ket{1}$ after normalization, with the corresponding LMP coefficient $D_R=1/\sqrt{2}$ compensating the branch success probability. Therefore, the measurement statistics with the LMP program gives the identical output as first applying the corresponding map and then performing the original measurements \cite{PhysRevLett.132.180201}. In this sense, the above transformation is absorbed into the measurement process, and the same statistics can be obtained through programmed local measurement bases and classical normalization, without implementing additional state-transformation modules as in Ref.~\cite{PhysRevLett.132.180201} and providing a more economical and experimentally accessible way to reproduce the desired virtual distillation effect. 
	
	More generally, the LMP shows that the desired distillation effect can be realized by selecting the proper local program. 
	By changing the programming measurement bases and normalization coefficients, different effective local CP maps and channels can be implemented.  
	This programmability allows the same framework to emulate local filtering operations, replacement channels, and other branch-wise CP transformations, indicating its flexibility for multipartite and higher-dimensional quantum systems.

	\begin{table}[h]
		\centering
		\renewcommand{\arraystretch}{1.3}
		\begin{tabular}{c cccccc|cccccc}
			\toprule\toprule
			
			& \multicolumn{6}{c|}{Alice} & \multicolumn{6}{c}{Bob} \\
			\midrule
			
			& $|0\rangle$ & $|1\rangle$ & $|+\rangle$ & $|-\rangle$ & $|R\rangle$ & $|L\rangle$ \ 
			& \ $|0\rangle$ & $|1\rangle$ & $|+\rangle$ & $|-\rangle$ & $|R\rangle$ & $|L\rangle$ \\
			\midrule
			
			$\Lambda_1$ & $|0\rangle$ & $|1\rangle$ & $|+\rangle$ & $|-\rangle$ & $|R\rangle$ & $|L\rangle$ & $|0\rangle$ & $|1\rangle$ & $|-\rangle$ & $|+\rangle$ & $|L\rangle$ & $|R\rangle$  \\
			
			$\Lambda_2$ & $|1\rangle$ & $|0\rangle$ & $|+\rangle$ & $|-\rangle$ & $|L\rangle$ & $|R\rangle$ & $|0\rangle$ & $|0\rangle$ & $|0\rangle$ & $|0\rangle$ & $|0\rangle$ & $|0\rangle$  \\
			$\Lambda_3$ & $|1\rangle$ & $|0\rangle$ & $|+\rangle$ & $|-\rangle$ & $|L\rangle$ & $|R\rangle$ & $|1\rangle$ & $|1\rangle$ & $|1\rangle$ & $|1\rangle$ & $|1\rangle$ & $|1\rangle$  \\
			
			\midrule\midrule
			
			& $D_0$ & $D_1$ & $D_{+}$ & $D_{-}$ & $D_{R}$ & $D_{L}$ \
			&\ $D_0$ & $D_1$ & $D_{+}$ & $D_{-}$ & $D_{R}$ & $D_{L}$ \\
			\midrule
			
			$\Lambda_1$ & 1 & 1 & 1 & 1 & 1 & 1 & 1 & 1 & 1 & 1 & 1 & 1 \\
			$\Lambda_2$ & 1 & 1 & 1 & 1 & 1 & 1 & $\sqrt{2}$ & 0 & $1/\sqrt{2}$ & $1/\sqrt{2}$ & $1/\sqrt{2}$ & $1/\sqrt{2}$  \\
			
			$\Lambda_3$ & 1 & 1 & 1 & 1 & 1 & 1 & 0 & $\sqrt{2}$ & $1/\sqrt{2}$ & $1/\sqrt{2}$ & $1/\sqrt{2}$ & $1/\sqrt{2}$  \\
			
			\bottomrule\bottomrule
		\end{tabular}
		
		\caption{Decomposition of the map $\sum_i \Lambda_i(\Psi^-)=\rho_\eta=(\mathbb{I}_4 - \ket{\Psi^-}\!\bra{\Psi^-})/3$, the corresponding operators and parameters.} \label{TA}
	\end{table}

	\section{Conclusions.}
	In this work, we propose and demonstrate the resource distillation through LMP protocols in both bipartite and tripartite cases. We observe the effect of purification through entanglement and diverse transformations of steering configurations enabled by the directional structures of quantum steering.  
	Our results indicate that the LMP framework exhibits good adaptability  across different quantum systems, while preserving scalability independent of the system size. This would pave the way for resource distillation in more robust and large-scale multipartite quantum networks \cite{PhysRevLett.78.3221,duan2001long,quantumnetwork1}.  
	
	Using the method outlined in Eq.~(\ref{nlf}), we further reformulate the virtual resource distillation as a local measurement program, enabling the required experimental apparatus to be replaced by a program applicable solely at the measurement process. This approach demonstrates the flexibility of our local measurement program, which can implement the desired quantum operation by appropriately redefining the measurement basis according to the structure of the corresponding completely positive map. Moreover, by effectively transferring the action of local CP maps onto the measurement settings, the LMP framework can be extended to other classes of local operations and more general CP maps like noise channels and POVM map\cite{RevModPhys.86.1203,Becerra2013POVM,BertlmannFriis2023QuantumChannels}. The LMP implementation provides a broadly applicable route for quantum resource manipulation \cite{scarani2009security,chiu2025continuous,sun2022optical}.
	
	\section{Acknowledgments}
	This work was supported by the National Natural Science Foundation of China Grants No. 12404403 and 123B2067, Zhejiang Provincial Natural Science Foundation of China Grant No.LQN25A040018, National Key Research and Development Program of China (Grant No. 2024YFA1408900), Hangzhou Leading Youth Innovation and Entrepreneurship Team project Grant No. TD2024005. 
	
	\section{Details of local program operation}
		For a local projective measurement $M_\phi=|\phi\rangle\langle\phi|$ and a CP operation $\mathcal{F}$, the transformed measurement effect can be written as
	\begin{equation}
		\mathcal{F}^\dagger M_\phi \mathcal{F}=r_\phi |\phi_\mathcal{F}\rangle\langle\phi_\mathcal{F}|,
	\end{equation}
	where
	\begin{equation}
		|\phi_\mathcal{F}\rangle
		=
		\frac{\mathcal{F}^\dagger|\phi\rangle}
		{\sqrt{\langle\phi|\mathcal{F}\mathcal{F}^\dagger|\phi\rangle}},\ 
		r_\phi=
		\langle\phi|\mathcal{F}\mathcal{F}^\dagger|\phi\rangle .
	\end{equation}
	Accordingly, the filtered measurement probability can be obtained from the original local state $\rho_L$ as
	\begin{equation}
		\mathrm{Tr}\!\left[M_\phi \rho_\mathcal{F}\right]
		=D_\mathcal{F}
		\mathrm{Tr}\!\left[
		|\phi_\mathcal{F}\rangle\langle\phi_\mathcal{F}|\rho_L
		\right],
		\label{projective_lmp}
	\end{equation}
	where $\rho_\mathcal{F}=\mathcal{F}\rho_L\mathcal{F}^\dagger/
	\mathrm{Tr}[\mathcal{F}\rho_L\mathcal{F}^\dagger]$ is the locally filtered state of the subsystem on which the local filter operation acts. 
	In practice, this means that the original measurement basis $|\phi\rangle$ is replaced by the updated basis $|\phi_\mathcal{F}\rangle$, while the corresponding outcome is renormalized by the normalization factor
	\begin{equation}
		D_\mathcal{F}
		=
		\frac{r_\phi}
		{\mathrm{Tr}[\mathcal{F}\rho_L\mathcal{F}^\dagger]} .
	\end{equation}
	
	
	In a practical experimental realization, taking the projective measurement as an illustrative example, the procedure can be implemented as follows: 
	(i) The user first characterizes the local reduced state $\rho_L$ of the subsystem, for example through local quantum state tomography. The local operation $\mathcal{F}$, which is chosen to enhance the desired resource correlations with other parties, can then be determined and further optimized through a feedback mechanism established from the measurement outcomes. 
	(ii) The user inputs $\mathcal{F}$ together with the original projective measurement basis $|\phi\rangle$ and the characterized local state $\rho_L$ into a programmable module. The module incorporates the completely positive transformation into the measurement process and generates the updated measurement basis $|\phi_\mathcal{F}\rangle$ together with the normalization factor $D_\mathcal{F}$. 
	(iii) The user then performs the measurement using the newly constructed basis $|\phi_\mathcal{F}\rangle$ and obtains the corresponding outcomes. These outcomes are multiplied by the normalization factor $D_\mathcal{F}$, yielding the final measurement results, which reproduce the effect of the completely positive map at the measurement level.
	
	As a result, the LMP substantially reduces the need for additional physical filtering elements and provides a flexible route for implementing resource distillation in multipartite and higher-dimensional quantum systems.
	
	\begin{figure*}[t]
		\includegraphics[width=0.7\textwidth]{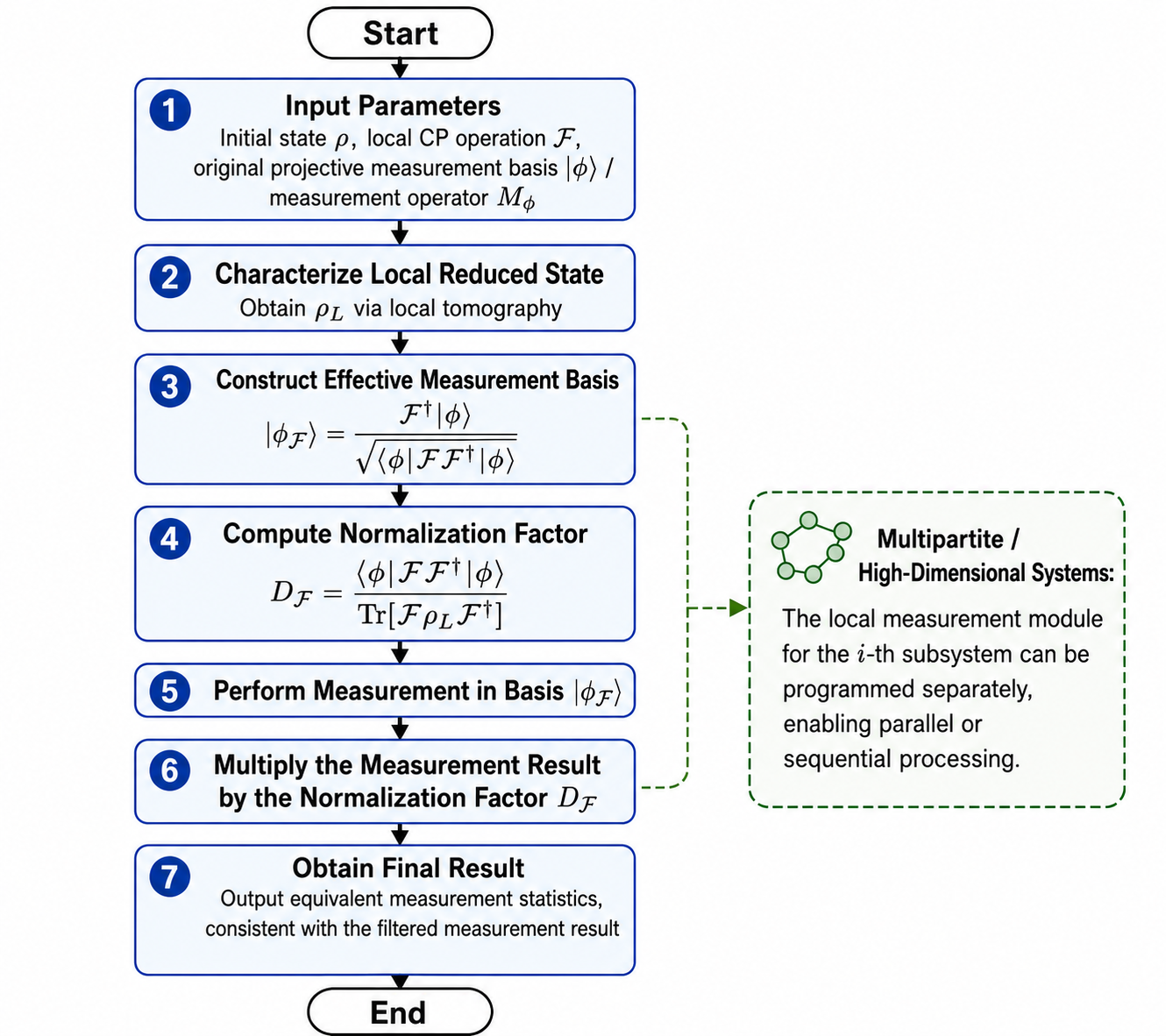}
		\caption{\textbf{The process of LMP in a practical experiment.} 
		}\label{sm}
	\end{figure*}
	
	\section{Criterion of quantum entanglement and steering in bipartite systems}
	In bipartite case, the degree of entanglement is quantified through concurrence $C=\max\{0,\Lambda=\lambda_1-\lambda_2-\lambda_3-\lambda_4\}$ with $\lambda_i$ corresponding to the decreasing eigenvalues of Hermitian matrix $\rho(\sigma_y\otimes\sigma_y)\rho^*(\sigma_y\otimes\sigma_y)$ \cite{PhysRevLett.80.2245}. 
	EPR steering criteria is based on local uncertainty relations in three-measurement settings $\{\vec{x},\vec{y},\vec{z}\}$, where Alice is able to steer Bob if the inequality 
	\begin{equation}
		\label{3}
		\sum_i\delta^2(\alpha_iA_i+B_i)\geq \mathrm{min}_{\rho_B}\sum_i\delta^2(B_i)
	\end{equation}
	is violated, with the variance of measurement outcomes $\delta$ and  $\alpha_i=-(\langle A_i B_i\rangle-\langle A_i\rangle \langle B_i\rangle)/\delta^2(A_i)$  \cite{PhysRevA.93.012108}. 
	Here, we use the steering parameter $ S_{ab}=2-\sum_i\delta^2(\alpha_iA_i+B_i)> 0$ demonstrates Alice could steer Bob, and $S_{ab}\le0$ means Alice can't steer Bob under three-setting measurements.

	\section{Criterion of quantum entanglement and steering in tripartite systems}
	As for the tripartite case, the entanglement witness $W=\frac{3}{4}\openone-P_{GHZ}$ is utilized with the projector $P_{GHZ}=\ket{GHZ}\bra{GHZ}$, $\ket{GHZ}=(\ket{000}+\ket{111})/\sqrt{2}$ and the identity matrix  $\openone$ \cite{PhysRevLett.87.040401}.  	The entanglement witness $W<0$ indicates the presence of entanglement. 
	
	The tripartite EPR steering is calculated in two different scenarios, i.e., the one-sided device-independent (1SDI) and two-sided device-independent (2SDI) cases \cite{cavalcanti2015detection}.   
	In 1SDI scenario, only Alice's measurement device is untrusted while Bob and Charlie are trusted, where the steerability from Alice to Bob and Charlie is denoted as $\mathcal{S}_{a|bc}=1+0.1547\langle Z_BZ_C\rangle-(\langle A_2Z_B\rangle+\langle A_2Z_C\rangle+\langle A_0X_BX_C\rangle-\langle A_0Y_BY_C\rangle-\langle A_1X_BY_C\rangle-\langle A_1Y_BX_C\rangle)/3$. 
	The steering parameter $\mathcal{S}_{a|bc}>0$ indicates that Alice can steer the joint state of Bob and Charlie. 
	In 2SDI scenario, Alice and Bob's measurement device are untrusted meanwhile Charlie are trusted, $\mathcal{S}_{ab|c}=1-0.1831(\langle A_2B_2\rangle+\langle A_2Z\rangle+\langle B_2Z\rangle)+0.2582(-\langle A_0B_0X\rangle+\langle A_0B_1Y\rangle+\langle A_1B_0Y\rangle+\langle A_1B_1X\rangle)$. 
	The steering parameter  $\mathcal{S}_{ab\rightarrow c}>0$ indicates that Alice and Bob can jointly steer Charlie’s state.


\end{document}